\documentclass[iop]{emulateapj} 
\usepackage{natbib} 
\usepackage{url} 
\usepackage{color} 
\newcommand{\etal}{et al.} 
\newcommand{\mic}{\mathrm{\mu s}} 
 
\slugcomment{version 3.5} 
\shorttitle{Correlation of \textit{Chandra} photons with the radio GPs from the Crab pulsar} 
\shortauthors{A.~V.~Bilous \etal} 

\begin{document} 
\title{Correlation of \textit{Chandra} photons with the radio giant pulses from the Crab pulsar} 
\author{A.~V.~Bilous\altaffilmark{1}, 
M.~A.~McLaughlin\altaffilmark{2,3},
V.~I.~Kondratiev\altaffilmark{4,5}, 
S.~M.~Ransom\altaffilmark{6}, 
} 
\altaffiltext{1}{Department of Astronomy, University of Virginia, PO Box 400325, Charlottesville, VA 22904; avb3k@virginia.edu} 
\altaffiltext{2}{Department of Physics, West Virginia University, Morgantown, WV 26506} 
\altaffiltext{3}{National Radio Astronomy Observatory, Green Bank, WV 24944} 
\altaffiltext{4}{Netherlands Institute for Radio Astronomy (ASTRON), Postbus 2, 7990 AA Dwingeloo, The Netherlands} 
\altaffiltext{5}{Astro Space Center of the Lebedev Physical Institute, Profsoyuznaya str. 84/32, Moscow 117997, Russia} 
\altaffiltext{6}{National Radio Astronomy Observatory, Charlottesville, VA 22903} 
\begin{abstract} 
No apparent correlation was found between giant pulses (GPs) and X-ray photons from 
the Crab pulsar during 5.4 hours of simultaneous observations with the Green Bank Telescope at 1.5\,GHz 
and \textit{Chandra X-Ray Observatory} primarily in the energy range 1.5$-$4.5\,keV. 
During the Crab pulsar periods with GPs the X-ray flux in radio emission phase windows does not change more than by $\pm10$\% for main pulse (MP) GPs and $\pm$30\% for interpulse (IP) GPs. During giant pulses themselves, the X-ray flux does not change  more than by two times for MP GPs and 5 times for IP GPs. All limits quoted are compatible with 2-$\sigma$ fluctuations of the X-ray flux around the sets of false GPs with random arrival times. The results speak in favor of changes in plasma coherence as the origin of GPs. However, the results do not rule out variations in the rate of particle creation if the particles that emit coherent radio emission are mostly at the lowest Landau level.  
\end{abstract} 


\section{Introduction} Giant pulses are a peculiar form of pulsar radio emission, clearly distinct from the 
regular pulses. Only a handful of pulsars have been reported to have GPs and neither the nature of GP 
emission nor the strict definition of a giant pulse are yet fully clear. The ``classical'' giant pulses, 
from the pulsars with high values of magnetic field at the light cylinder, are short (ns -- $\mu$s), bright 
(up to MJy) and obey a power-law energy distribution \citep[see ][ and references therein]{kuzmin2007}. These 
properties obviously point to the coherent nature of the emission mechanism. 

It is interesting that for the 
pulsars  with ``classical`` GPs and pulsed X-rays or $\gamma$-rays, giant pulses come from the same pulse 
phase range as high-energy emission \citep{knight2006_china, cusumano2003}. That indicates that GPs and high-energy 
photons may originate in the same region in the magnetosphere. Without any doubt, the fact that GPs are (or are not) 
accompanied by increasing high-energy flux is important for testing the theories of GP generation. On a broad scale, 
if the GP emission mechanism involves some change in particle creation rate or beam direction alteration, one should 
observe a correlation between GPs and high-energy emission. If GPs are due to  changes in coherence in 
magnetospheric plasma, there should be no correlation, since high-energy emission is of incoherent origin. 
Also, while propagating in the magnetosphere, radio photons can be shifted to a higher energy band by interacting with the relativistic plasma via inverse Compton scattering or resonant absorption. This would result in potentially detectable radio/high-energy correlation.

\begin{table*}[thb]
\begin{center} 
\caption{Previous studies of correlation between GPs from the Crab pulsar and its high-energy emission\label{table_sum}}
%\begin{tabular}{p{3cm}|p{2.6cm}|p{2.6cm}|p{2.6cm}|p{2.6cm}|p{2.6cm}} 
\begin{tabular}{lccccc} 
\hline
& \parbox[c]{2.3cm}{\bf $\frac{{}^{{}^{}}}{{}^{{}^{}}}$\citet{shearer2003}} &
%\parbox{2.3cm}{\bf Lundgren et al. \\ (1995)} &
\parbox{2.3cm}{\bf \citet{lundgren1995}} &
\parbox{3.0cm}{\bf \citet{ramanamurthy1998}} & 
\parbox{2.3cm}{\bf \citet{bilous2011}} & 
\parbox{2.6cm}{\bf Mickaliger et al. \\ (in prep)} \\ 
\hline\smallskip
\parbox{2.6cm}{High-energy observation energy range }& 6000$-$7500\,$\AA{}$ & 50$-$220\,MeV & 0.05$-$30\,GeV & 0.1$-$5\,GeV & 0.1$-$300\,GeV \\ [0.1cm]
\parbox{2.6cm}{Radio observation central frequency (GHz)} & 1.4 & 1.3, 0.8 & 1.3, 0.8 & 8.9 & 1.2, 0.33 \\ [0.2cm]

\parbox{2.6cm}{Radio observation bandwidth (MHz)} & 5 & 20 & 20 & 800 & 400, 150 \\ [0.2cm]

\parbox{2.6cm}{GP selection threshold (Jy)} & 150 (peak) & 125 (mean) & 125 (mean) & 8.1 (peak) & S/N$>$10 \\ [0.2cm]
Observing time (h) & 3 & 14  & 32 & 10 & 30 \\ [0.2cm]
\parbox[t]{2.6cm}{High-energy flux change in units of average flux during observations${}_{{}_{}}$} & 
\parbox[t]{2.6cm}{3\% enhanced flux (7.8$\sigma$) in 0.02 phase window around MP GPs} & 
\parbox[t]{2.6cm}{with a 95\% confidence level, $<$2.5, in a period around GPs and up to $\pm 5$ periods} & 
\parbox[t]{2.6cm}{with a 99.9\% confidence level, $<$4.6 for 100\,ms window around GP} & 
\parbox[t]{2.6cm}{with a 95\% confidence level, $<$4 for 30\,ms around GPs, $<$12 for 0.2 phase window around IP GPs} & 
\parbox[t]{2.6cm}{maximum 1$\sigma$ correlation in one period around GPs} \\
\hline 
\end{tabular} 
\end{center}
\end{table*}

The Crab pulsar is the most studied pulsar with GPs and the easiest object for such correlation studies, since 
the pulsar is relatively bright at all wavelengths. GPs from the Crab appear in phase regions of the main pulse 
(MP) and interpulse (IP) at all radio frequencies where the Crab has been observed (from 23\,MHz up to 15\,GHz). Also, the detection of GPs in high frequency components at 8.35 GHz has been reported by  \cite{jessner2005}, albeit not confirmed by later observations \citep{jessner2010}.
 
Around 4\,GHz the interpulse disappears and then reappears shifted ahead by about 10 degrees \citep{moffett1996}. Curiously, above 4\,GHz IP GPs have totally different spectral properties than MP GPs at the same frequency \citep{hankins2007}. Thus, IP GPs above 4\,GHz may be produced by a separate emission mechanism. 

Previously, there have been several correlation studies between Crab GPs and its high-energy emission 
(see Table~\ref{table_sum} for summary). The only positive results so far were obtained in the optical light by 
\citet{shearer2003}. They found that optical pulses coincident with GPs were on average 3\% brighter than the others. A number of $\gamma$-ray studies covering the energy range from 50\,MeV to 300\,GeV did not find any statistically significant correlation, showing that the $\gamma$-ray flux around GPs does not change by more than a factor of several times (\citealt{lundgren1995}, \citealt{ramanamurthy1998}, \citealt{bilous2011}, Mickaliger et al., in prep).

A few studies have also been conducted of Crab X-ray variability based on high-energy data alone. The analysis 
of 1 hour of RXTE data (1$-$10\,keV) showed that the flux from the Crab pulsar is stable at the level of 7\% \citep{pat1999}. 
Also, in the 13.3 $-$ 58.4\,keV, based on 18 hours of RXTE data, \citet{vivekanand2001} found the rms of X-ray flux variation in 
time at any phase in the on-pulse window to be within of $\approx$3.5\% of the corresponding mean flux of the integrated 
profile of Crab pulsar. 

These studies ruled out the existence of short strong ``GP-like" bursts of X-ray emission, but 
they were not capable of detecting small sporadic X-ray flux variations, as could be expected around GPs. In this paper we 
used 5.4 hours of simultaneous observations of the Crab pulsar with the 100-m Robert C. Byrd Green Bank Telescope (GBT) and the \textit{Chandra X-ray Observatory} to perform an extensive correlation of GPs with X-ray photons.

\section{Radio observations} 
\label{sec:radioobs}

\begin{figure} 
\includegraphics[scale=0.45]{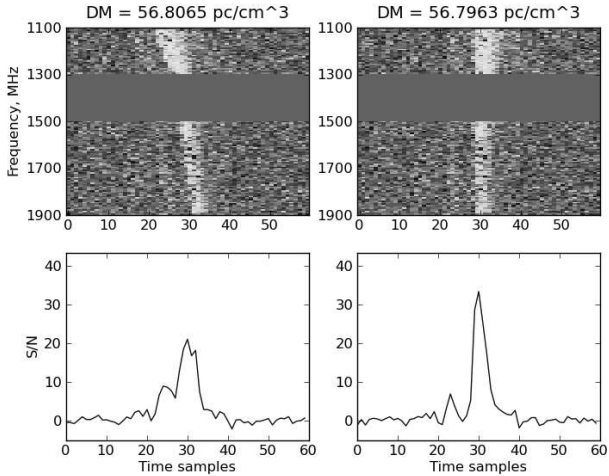}
\caption{An example of a strong GP, dedispersed with the $\mathrm{DM}=56.8065$ pc cm${}^{-3}$ from the Jodrell Bank Crab Pulsar Monthly Ephemeris from Nov 15, 2010 (left) and with the correct DM, corrected by 0.0102\,pc\,cm$^{-3}$ (right). The gap between 1.3 and 1.5 GHz is due to a backend malfunction. \label{fig:dms}} 
\end{figure}

\begin{figure*} 
\includegraphics[scale=0.45, angle=0]{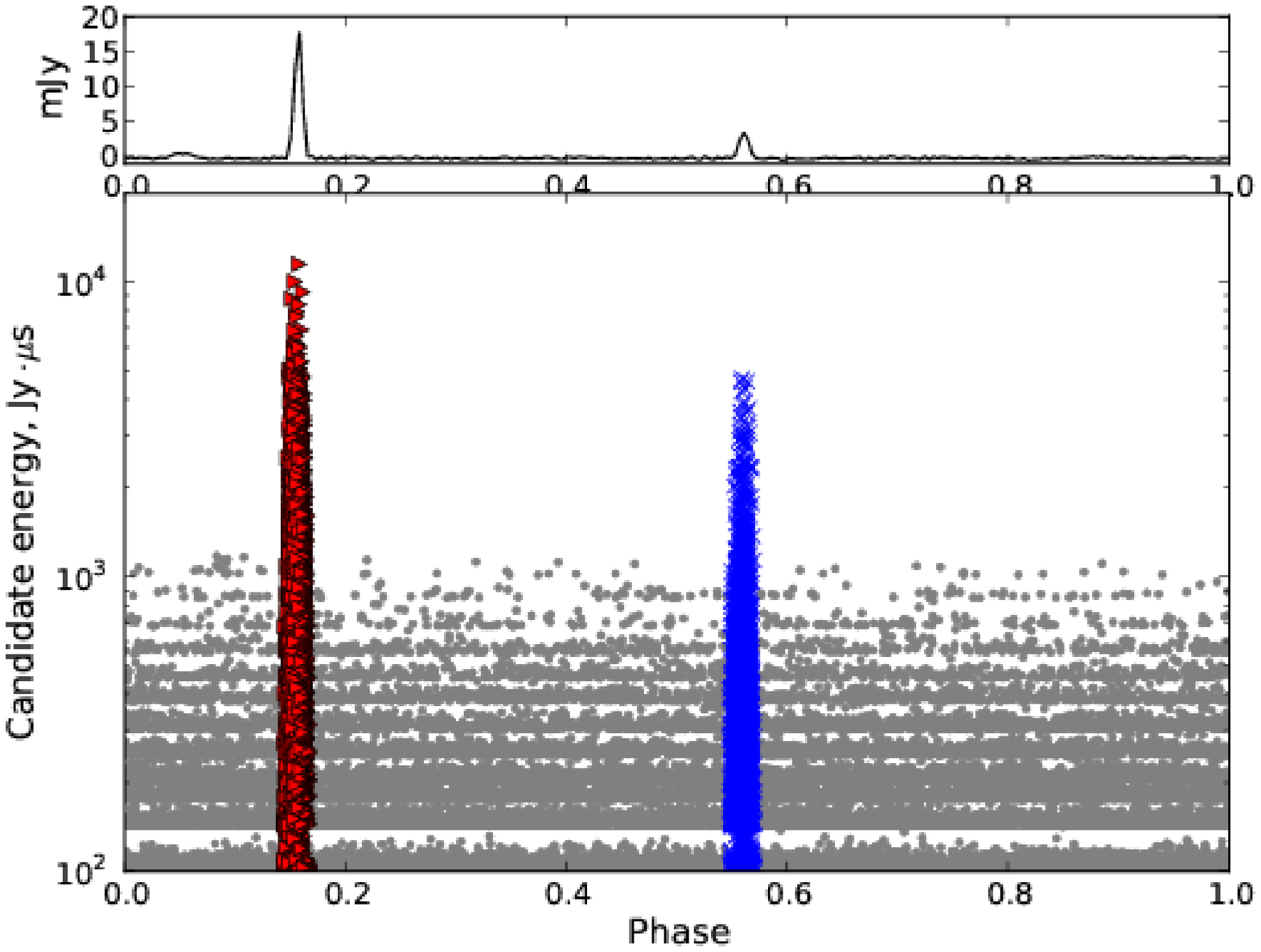}\includegraphics[scale=0.45]{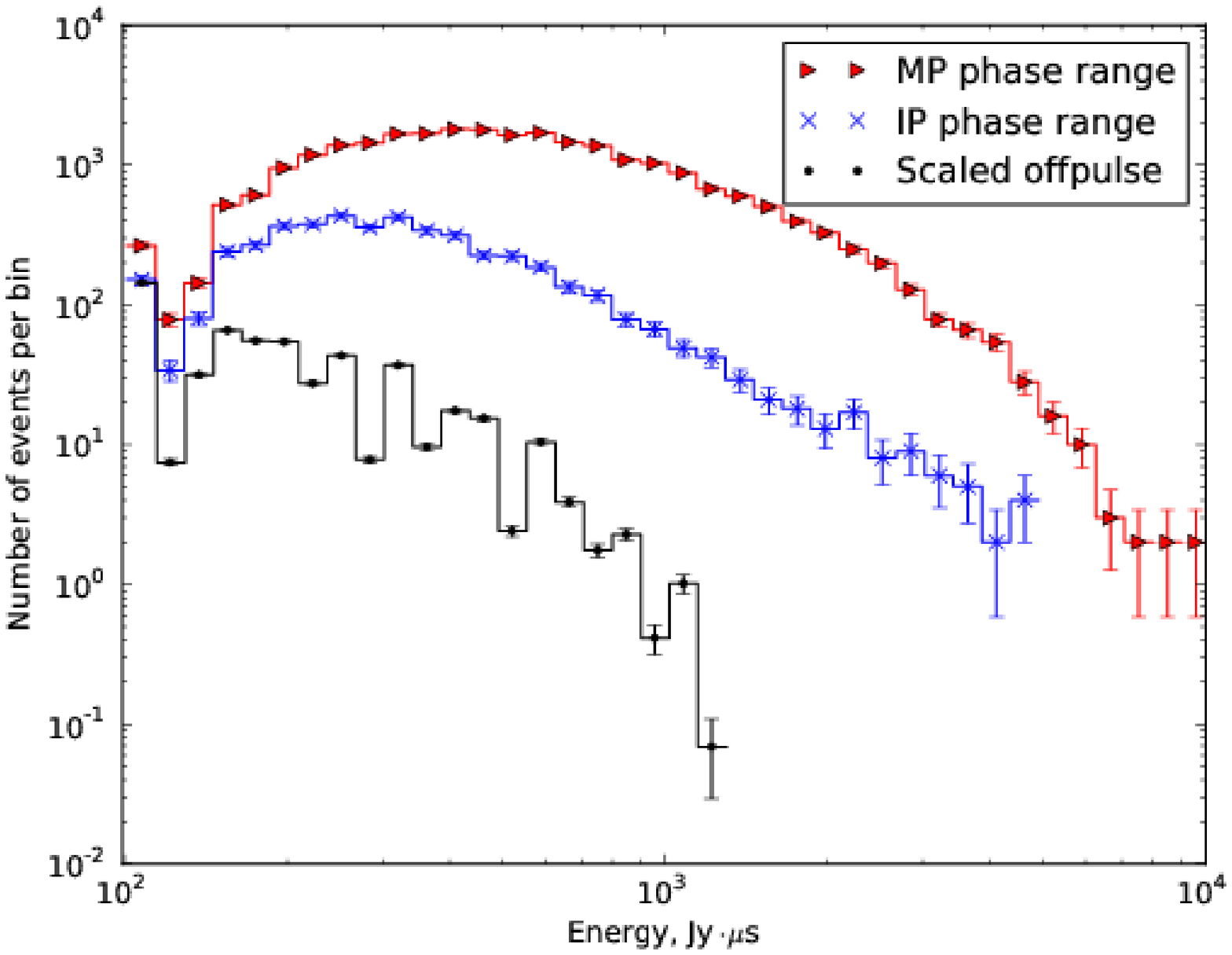} 
\caption{\textit{Left:} GP candidates, selected by matched-filtering with a set of boxcars of different width. Most of the candidates come from phase regions coincident with the MP (red triangles) and IP (blue crosses). Events outside the MP/IP (grey dots) tend to concentrate at a set of levels because of the width-dependent selection thresholds (see text for explanation). The average profile from the same observation is plotted on top for reference. \textit{Right:} Energy distributions of selected GPs and spurious detections. The energy distribution of spurious detections was scaled by 0.023 to match the size of MP/IP phase windows.  The shape of the distributions below 1011\,Jy$\cdot\mic$ is affected by width-dependent selection thresholds.  \label{fig:e_distr}} 
\end{figure*} 

Radio observations were carried out on 2010 November 16 with the GBT using the Green Bank Ultimate Pulsar Processing Instrument (GUPPI) at a central frequency of 1.5\,GHz in coherent dedispersion search mode. The total nominal band of 800 MHz was split into 128 frequency channels and the full Stokes parameters were recorded with a sampling interval of $t_{\mathrm{res}}=2.56$\,$\mic$. Unfortunately, the data from 1.3 to 1.5\,GHz were lost because of a backend malfunction, thus reducing the effective bandwidth to 600\,MHz. Radio observing time comprised 5.4 hours or $5.8\times10^5$ pulsar periods. 

At 1.5\,GHz the system temperature is dominated by the Crab nebula, which is smaller \citep{green2009} than the $8.2\arcmin$ GBT beam at this frequency. We calculated the flux density for the Crab Nebula with the relation $S(f) = 955\times(f/\mathrm{GHz})^{-0.27}$\,Jy \citep{cordes2004}. With the receiver temperature $T_\mathrm{rec}=20$\,K, background sky temperature $T_\mathrm{bkg}=1.5$\,K, and gain $G=2$\,K$/$Jy, our estimated system equivalent flux density SEFD~$= 7.9$\,Jy for a 600\,MHz bandwidth and a sampling time of 2.56\,$\mic$. 

During the observations, we performed coherent dedispersion in each of the frequency channels with the most up-to-date value of dispersion measure (DM) from the Jodrell Bank Crab pulsar monthly ephemeris\footnote{\url{http://www.jb.man.ac.uk/$\sim$pulsar/crab.html}, see also \citet{lyne1993}}, $\mathrm{DM_{JB}}=56.8065$\,pc\,cm$^{-3}$. However, the dispersion measure of the Crab pulsar is known to vary on  monthly timescales by $\sim 0.01$\,pc\,cm$^{-3}$. Such error in DM will not result in detectable pulse smearing within one frequency channel, but across 800\,MHz bandwidth the net smearing can reach 20\,$\mic$, or 8 time samples. To check the DM, we looked at the spectra of the brightest events dedispersed with $\mathrm{DM_{JB}}$. All events exhibited the characteristic quadratic trend across the band, as expected for an error in DM (see Fig.~\ref{fig:dms}, left). We dedispersed these events with a set of trial DM values and for each pulse found the DM which corresponded to maximum peak S/N. The mean of these DM values was 56.7963\,pc\,cm$^{-3}$ with standard deviation of 0.001\,pc\,cm$^{-3}$. An example of a giant pulse dedispersed with the corrected dispersion measure is shown in Fig.~\ref{fig:dms} (right). 

After correcting the dispersion measure, we dedispersed the raw data using the PRESTO package\footnote{\url{http://www.cv.nrao.edu/$\sim$sransom/presto/}}. For the correlation study, we used only the total intensity time series. More detailed investigation of the recorded GPs, including spectra, polarization, width and energy distribution will be presented elsewhere. 

To search for GPs we convolved the dedispersed time series with a series of boxcar functions of varying boxcar width. We used \texttt{singlepulse\_search$.$py} from PRESTO, which accounts for all the possible ``phase`` offsets of each boxcar and records all events with S/N$>$5 on the \textit{averaged} signal. Such definition of the selection threshold facilitates sifting the duplicate candidates (the ones above the threshold for different boxcar widths, but with the same time of arrival). While comparing  different boxcar widths, the peak S/N of the averaged signal has a clear maximum when the boxcar width matches the pulse width.  However, selecting a candidate by its S/N on the \textit{averaged} signal effectively results in a width-dependent energy threshold. After averaging by $n$ samples, the standard deviation of the signal drops by $\sqrt n$. Therefore S/N of 5 on the averaged signal corresponds to a mean S/N of $5/\sqrt n $ on the signal with original time resolution. The energy of each candidate was calculated by multiplying the mean S/N by the corresponding boxcar width. Thus, the minimum detectable energy of the $n$-sample wide candidate was $5\sqrt n\cdot\mathrm{SEFD}\cdot t_{\mathrm{res}}$, or 101.1$\sqrt{n}$\,Jy$\cdot\mic$ with SEFD of 7.9\,Jy and time resolution of 2.56\,$\mic$. In our analysis we used boxcars of 1, 2, 3, 4, 6, 9, 14, 20, 30, 45, 70 and 100 samples and the width-dependent energy thresholds can be seen on Fig.~\ref{fig:e_distr} (left) as the concentration of events at the energy levels proportional to the square root of the boxcar widths used. 

 Most of GP candidates were detected within two narrow  phase regions, each spanning 0.022 pulsar rotations and coincident with the main pulse and the interpulse on the average profile, obtained by folding the dedispersed time series with the pulsar period (see Fig.~\ref{fig:e_distr}, left). Outside the MP/IP region the phase distribution of events is flat, as it should be for noise. No excess of events was found in the Low Frequency Component (peaks at the phase 0.05). 

 As has been reported by \citet{popov2007}, giant pulses from the MP and IP have different energy distributions: IP GPs are well fitted by a single power-law, whereas MP GPs obey a broken power-law, with the energy of the break depending on observing frequency as $E_{\mathrm{break}} = 7\left(\nu/\mathrm{GHz}\right)^{-3.4}$\,kJy$\cdot\mic$. For our observing frequencies, $E_{\mathrm{break}}$ ranges from 0.8 to 5.0\,kJy$\cdot\mic$, making the distribution of MP GPs detected in the 800\,MHz band a convolution of several different broken power-laws. 

\begin{figure*} 
\includegraphics[scale=0.35]{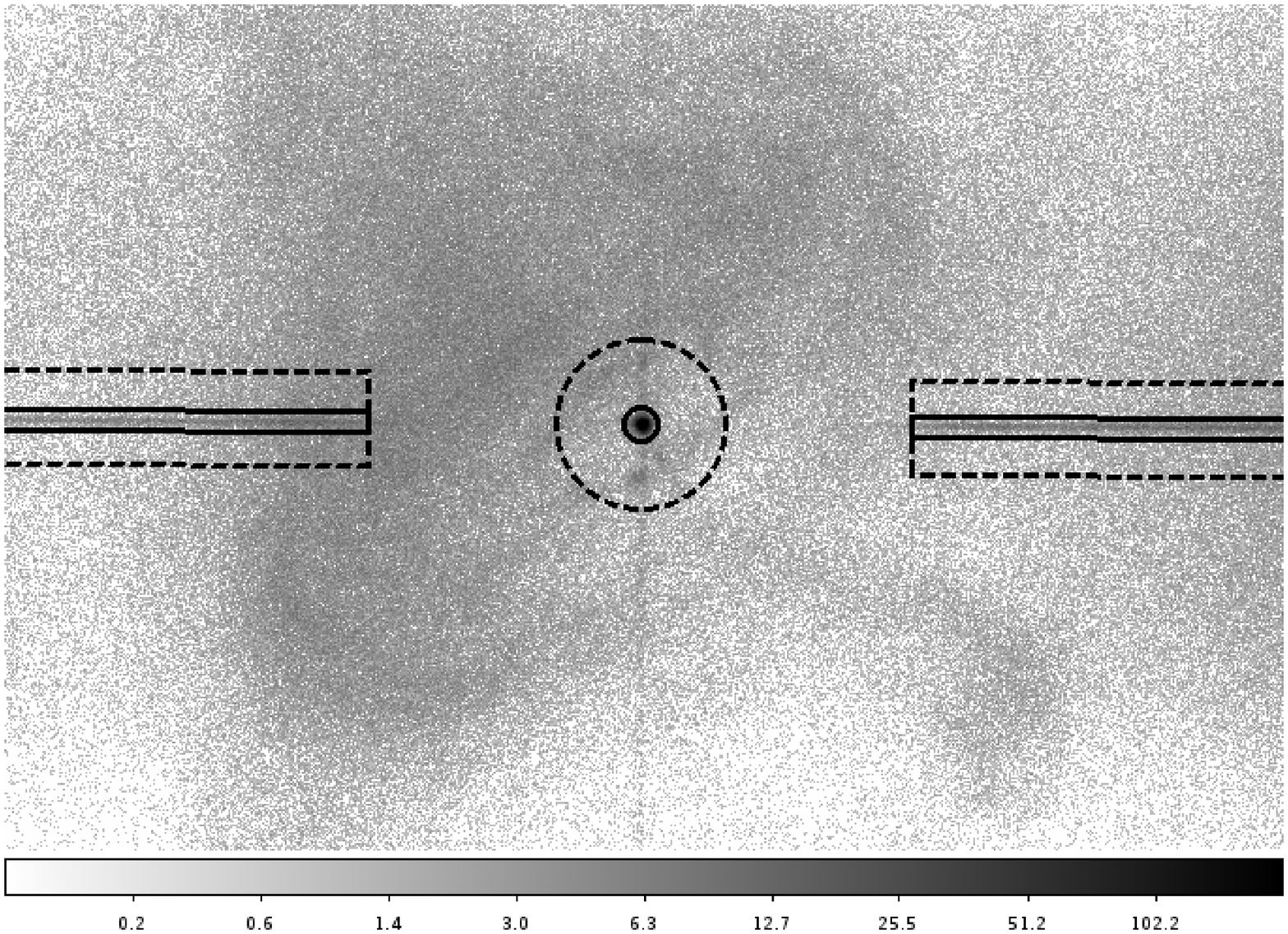}\includegraphics[scale=0.45]{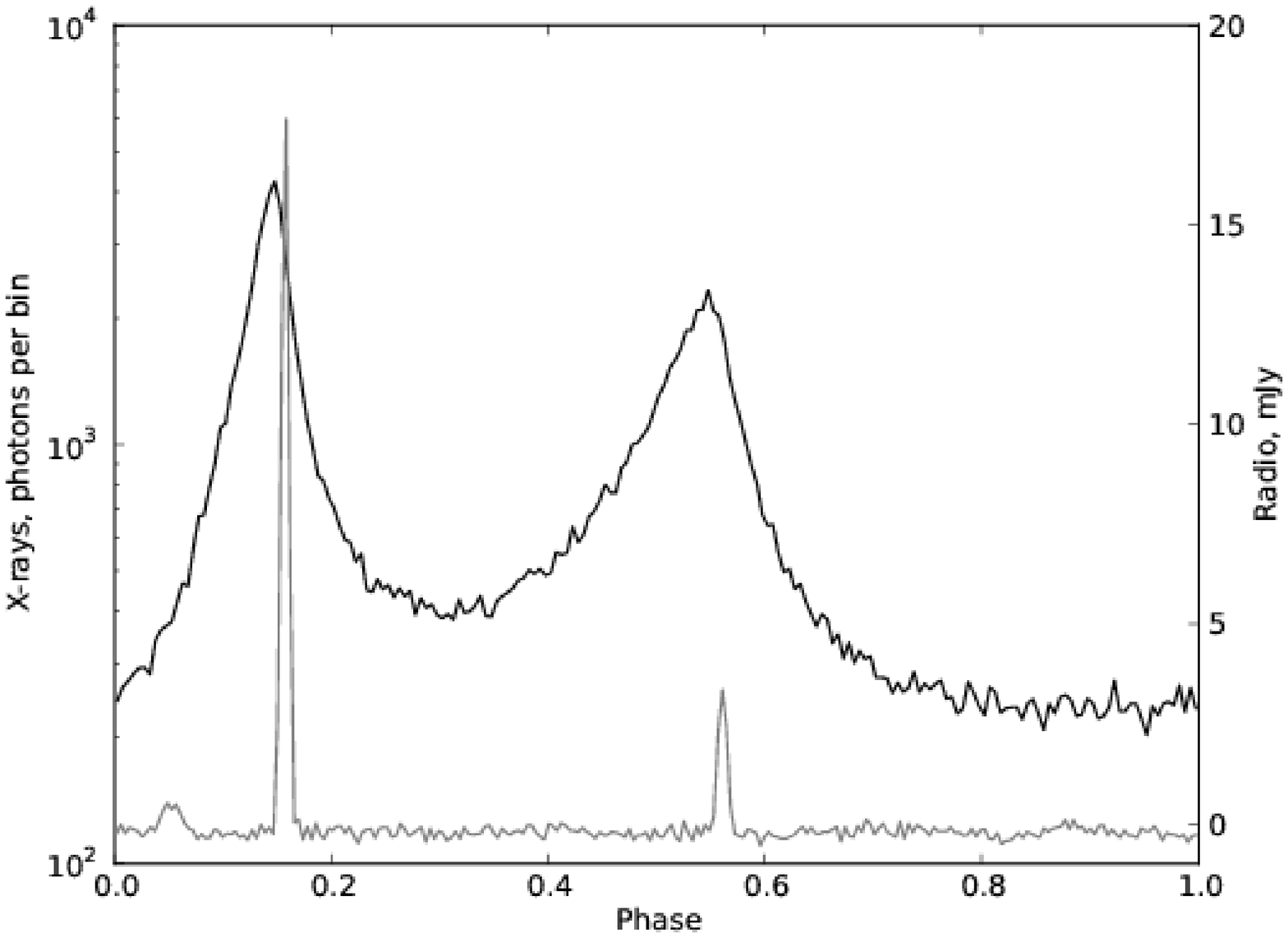} 
\caption{\textit{Left:} central part of the HRC-S image. Solid line: source region, zero order (circle) and dispersed spectra (boxes, only central part shown). Dashed line: background region. Due to an imperfect PSF model, we still detect weak pulsed emission in the background, so we used only photons within pulsar phases 0.7$-$1.0. \textit{Right:} phase distribution of the extracted X-ray photons (black). The mean profile of the radio emission is overplotted for reference (grey).\label{fig:xray_reg}} 
\end{figure*} 

As can be seen from Fig.~\ref{fig:e_distr} (right), above 300\,Jy$\cdot\mic$ the distributions for events in MP and IP windows behave as predicted. We believe that the roll-off below 300\,Jy$\cdot\mic$ is most probably due to the underestimation of width (and thus energy) for faint pulses.  Also, all distributions are affected by width-dependent selection thresholds. This impact is best visible as a dip between 101.1\,Jy$\cdot\mic$ and 101.1$\sqrt{2}$\,Jy$\cdot\mic$. In any case, the lowest energy of events is 20 (for MP window) or 100 (for IP) times bigger than the corresponding energy of the mean profile. By the formal definition of GPs as those with energy exceeding 10 times the energy of the mean profile \citep{Cairns2004}, even the faintest pulses in our sample can be considered giant. Thus, for the subsequent analysis we kept all events in the on-pulse phase range, 25986 in MP and 4636 in IP. In the off-pulse phase range, which spanned 0.956 pulsar rotations, 23366 events were detected. Thus, we estimate the fraction of \textit{false GPs} to be about 2.1\% in the MP window and 11.6\% in the IP window. 

We put the list of GP arrival times in TEMPO2 format \citep{Tempo2} and converted to the barycentric reference frame for the subsequent correlation analysis with \textit{Chandra} data. We used the TDB time system and DE405 solar system ephemeris.  Times of arrival (TOAs) were corrected for delay due to propagation in the ionized interstellar medium (ISM). Timing errors due to errors in DM are up to 1.8\,$\mic$. 

The observed rate and peak flux densities of GPs are inevitably affected by scintillation in the ISM and the Crab Nebula. To find the scintillation timescales we constructed the autocorrelation function for the GP rate. We did not find any GP rate change on any timescales down to 1\,s, the smallest bin size possible with our limited sample of GPs. Also, we found no scintillation patterns in the two-dimensional (observing frequency and time) autocorrelation function for the dedispersed data \citep{rickett1977}, averaged over 5 seconds to increase the signal to noise ratio.

By fitting an exponential scattering tail to the average profile of the strongest pulses (S/N$>40$), we estimated the scattering time to be about 10.3\,$\mic$, which is about 10 times bigger than the expected broadening by interstellar scattering \citep{Kuzmin2002} and could be explained by scintillations of the nebula or by the intrinsic duration of GPs \citep{Crossley2010}.  In any case, both expected and measured scattering times give a decorrelation bandwidth at least an order of magnitude smaller than the width of one frequency channel. The refractive interstellar scintillation (RISS) timescale, $\tau_{\mathrm{RISS}}$, can be scaled using the $\nu^{-2.2}$ dependence derived from a five frequency dataset \citep{rickett1990}. For 1.5\,GHz it corresponds to $\tau_{\mathrm{RISS}}$=1.2\,days, much bigger than our observing timespan. The diffractive interstellar timescale is $\tau_{\mathrm{DISS}} = \tau_{\mathrm{RISS}}\cdot \triangle f_{\mathrm{DISS}}/f = 1$\,s. However, we did not have enough GPs or sensitivity in folded data series to detect scintillations on such short timescales. 

\section{X-ray observations} 
\label{sec:xrayobs} 

The X-ray observations were carried out with \textit{Chandra}, using the Low-Energy Transmission Grating (LETG) and High-Resolution Camera spectroscopy detector (HRC-S) in timing mode (ObsId 11245). To ensure good timing accuracy we used a specific HRC set-up that selected only signals from the central HRC-S segment, disabled all on-board vetoing, and set the trigger-level = 50. The errors in photon arrival times in timing mode are about 16\,$\mic$\footnote{\url{http://asc.harvard.edu/cal/Hrc/}}. Simultaneous observing time with the GBT comprised 5.4 hours.

We created the level=2 photon lists using the data from the \textit{Chandra} X-ray Center pipeline\footnote{\url{http://cxc.harvard.edu/ciao/guides/hrc$\_$data.html}}, and barycentered the arrival times with \texttt{axbary} from the CIAO software package. \texttt{Axbary} converts all arrival times to the TDB time system, so we could assign each photon its pulse phase using TEMPO2 with the same pulsar and solar system ephemeris as for the radio data. 

\begin{figure*} 
\includegraphics[scale=0.45]{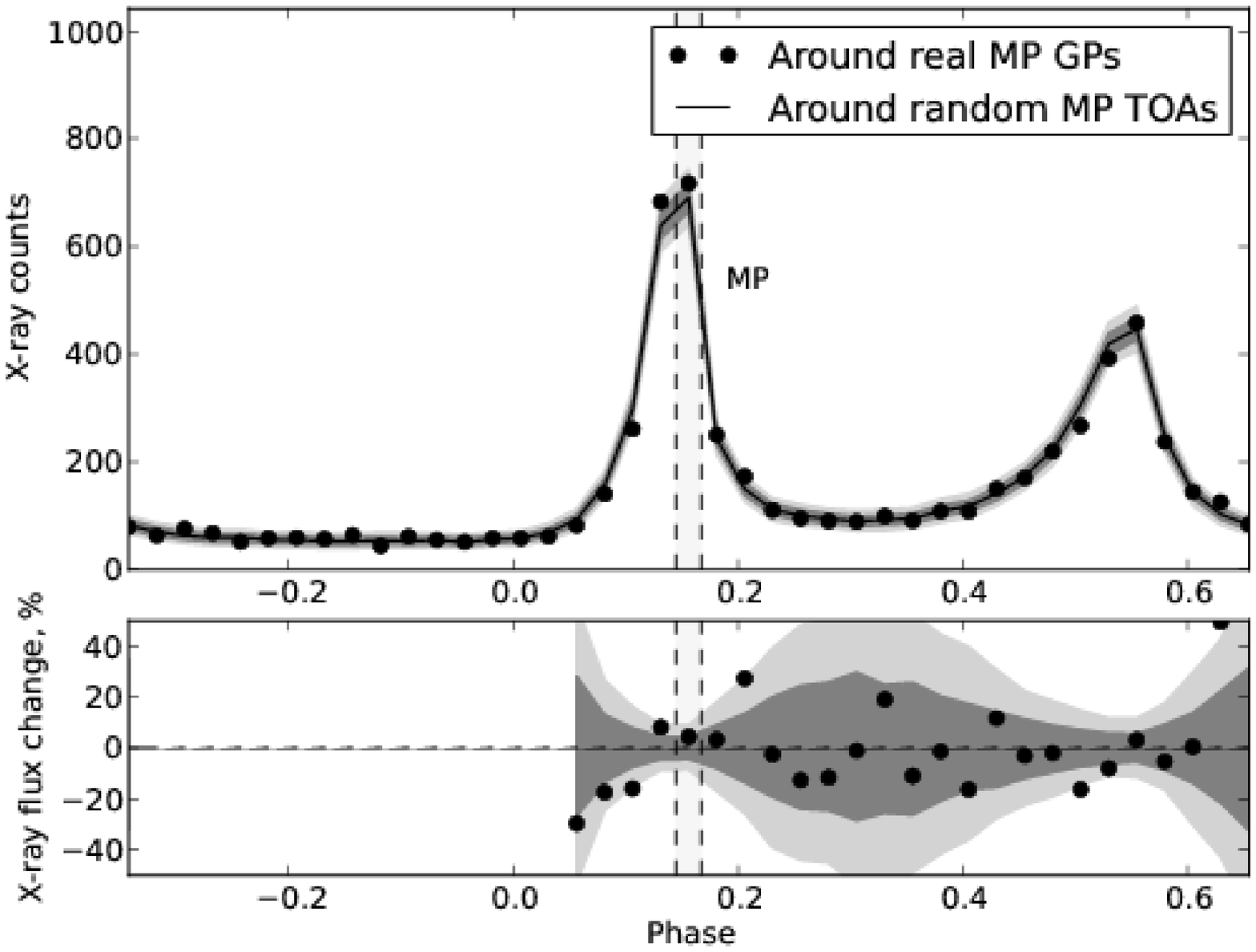}\includegraphics[scale=0.45]{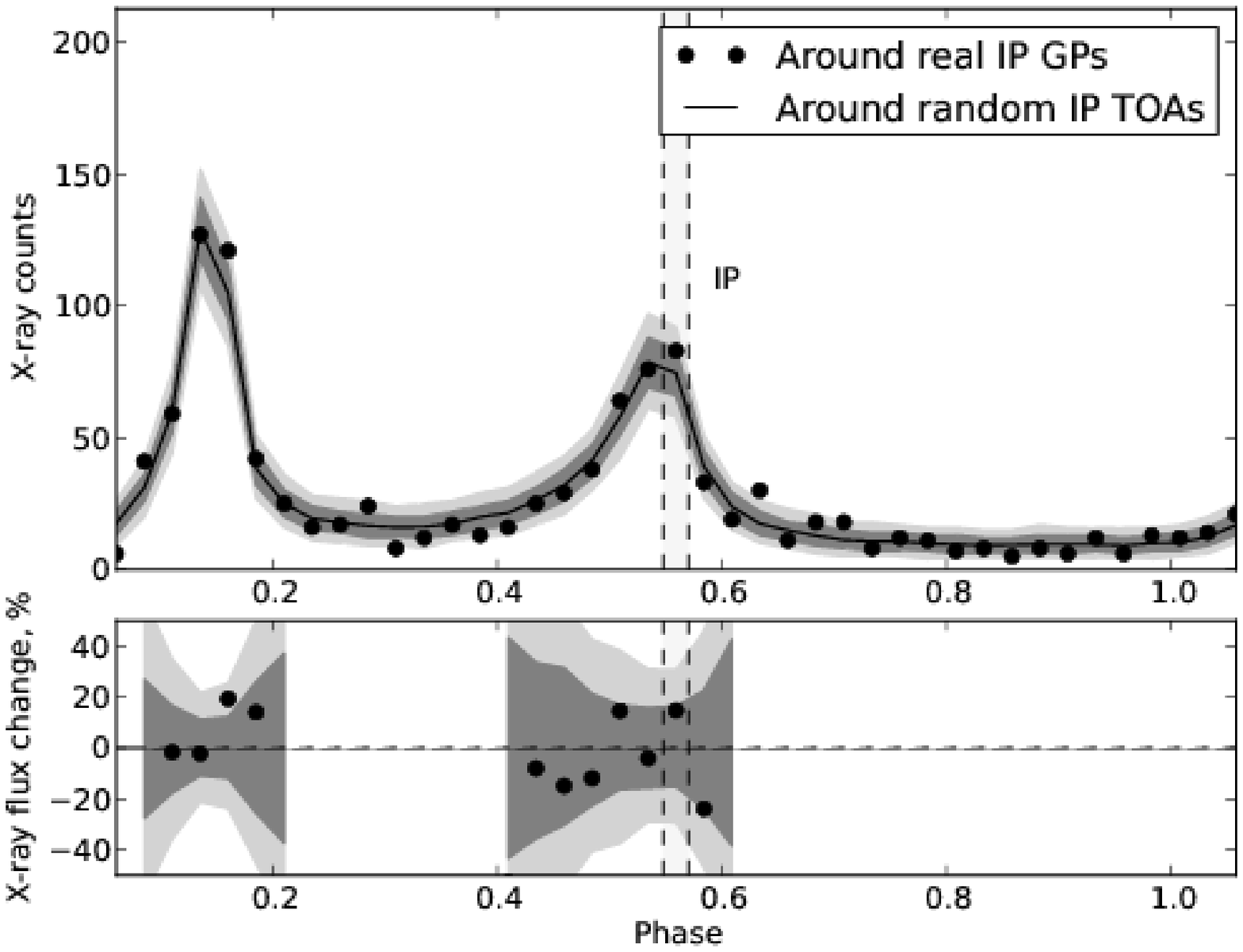} 
\caption{Upper subplots: X-ray counts in the radio emission region around MP GPs (left, within dashed lines) and IP GPs (right, within dashed lines) together with X-ray counts in the windows of similar size, shifted by the size of the window within one pulsar period from a GP. Black line: X-ray counts around the same number of MP/IP GPs with randomized integer part of phase. Shaded regions -- 68\% and 95\% percentile on the pool of simulated datasets.  Bottom subplots: relative change of X-ray flux $(f_{\mathrm{GP}}-f_{\mathrm{noGP}})/f_{\mathrm{noGP}}$. Only regions with good statistics are shown.\label{fig:corr_e}} 
\end{figure*}

While checking the X-ray data, we noticed that the count flux from any region on the detector gradually grows by about 20\% over the course of observations. Specifically, if $f(t)$ is the count flux in some region at a certain moment, then $f(t)$ can be expressed as $k(t)\cdot\overline{f(t)}$, where $\overline{f(t)}$ is the average count flux in that region and $k(t)$ has roughly the same shape for all regions. Interestingly, the phase-resolved flux exhibits the same behavior -- the increase of flux in each phase bin is proportional to the mean value of flux in that phase bin. Such a variability pattern cannot be explained by increasing background, since its change would be independent from the pulsar phase.
 Discussions with the HRC instrument engineers and scientists did not result in an explanation for the rate change (CXC Help Desk Ticket \#13509).  The detector particle background appeared to be essentially flat throughout the observation. The only change in instrument conditions seen was a slight temperature change. However, we have no mechanism to explain how it could change the X-ray rate but not the charged-particle rate (Michael Juda, private communication).

Since X-ray flux changes gradually and GP rate is constant over the course of observations, we can still constrain the change in X-ray flux around GPs, by comparing the X-ray flux around GPs to the X-ray flux around random time points with the same pulsar spin phase range as for the GPs.  However, measuring the absolute value of the flux is currently not possible.

For the subsequent analysis we extracted photons from $1.9\arcsec$ radius circle around the pulsar image in the 0th order LETG spectrum and from the two boxes around higher spectral orders (see Fig.~\ref{fig:xray_reg}, left, solid line). We did not make any cuts by energy, but most photons in our sample had energies of 1.5$-$4.5\,keV. Figure~\ref{fig:xray_reg}, right shows the phase distribution for our X-ray sample, with the radio profile overplotted for reference. Giant pulses (which coincide with the peak of radio emission) are delayed with respect to the maximum of the X-ray profile by about 0.01 of pulsar phase. This confirms the result reported by \citet{rots2004}. 

The average count rate on the whole detector was about 100~cts/s, which is well below telemetry saturation limit of 184~cts/s. The average dead time fraction was 0.9897, so the dead time after each recorded photon comprised 100\,$\mic$. This time is bigger than the duration of a typical GP and only a few times smaller than the size of MP/IP emission window. However, since the average number of X-ray photons per one pulsar period in our observations is quite low -- only 3 photons from the whole detector and  0.2 photons from our region of interest, the dead time will not considerably affect the correlation results on any time scale, unless the correlated photons come in bunches of less than 100\,$\mic$.

\section{Correlation Analysis} 
\label{sec:corr} 

\begin{figure*} 
\includegraphics[scale=0.45]{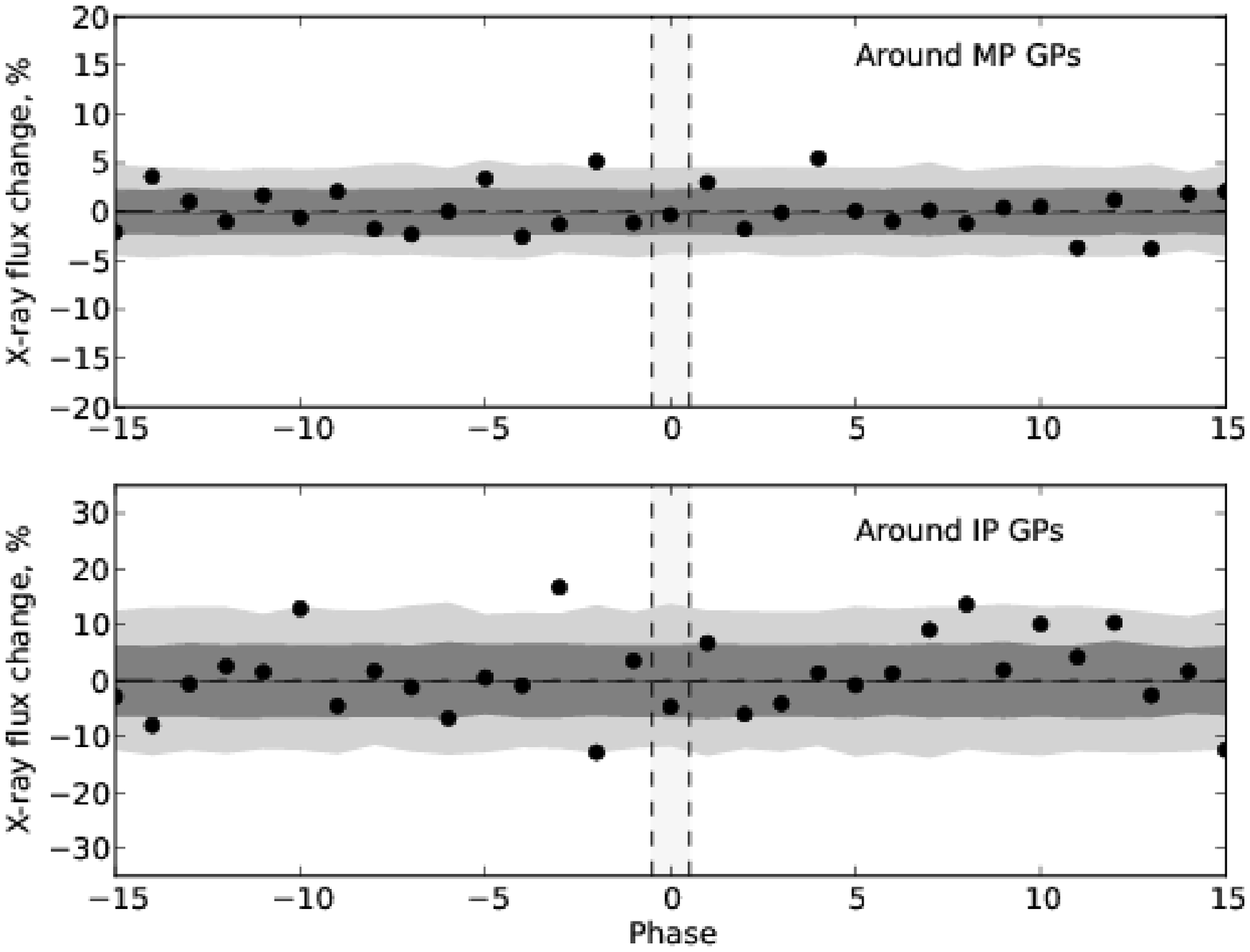}\includegraphics[scale=0.45]{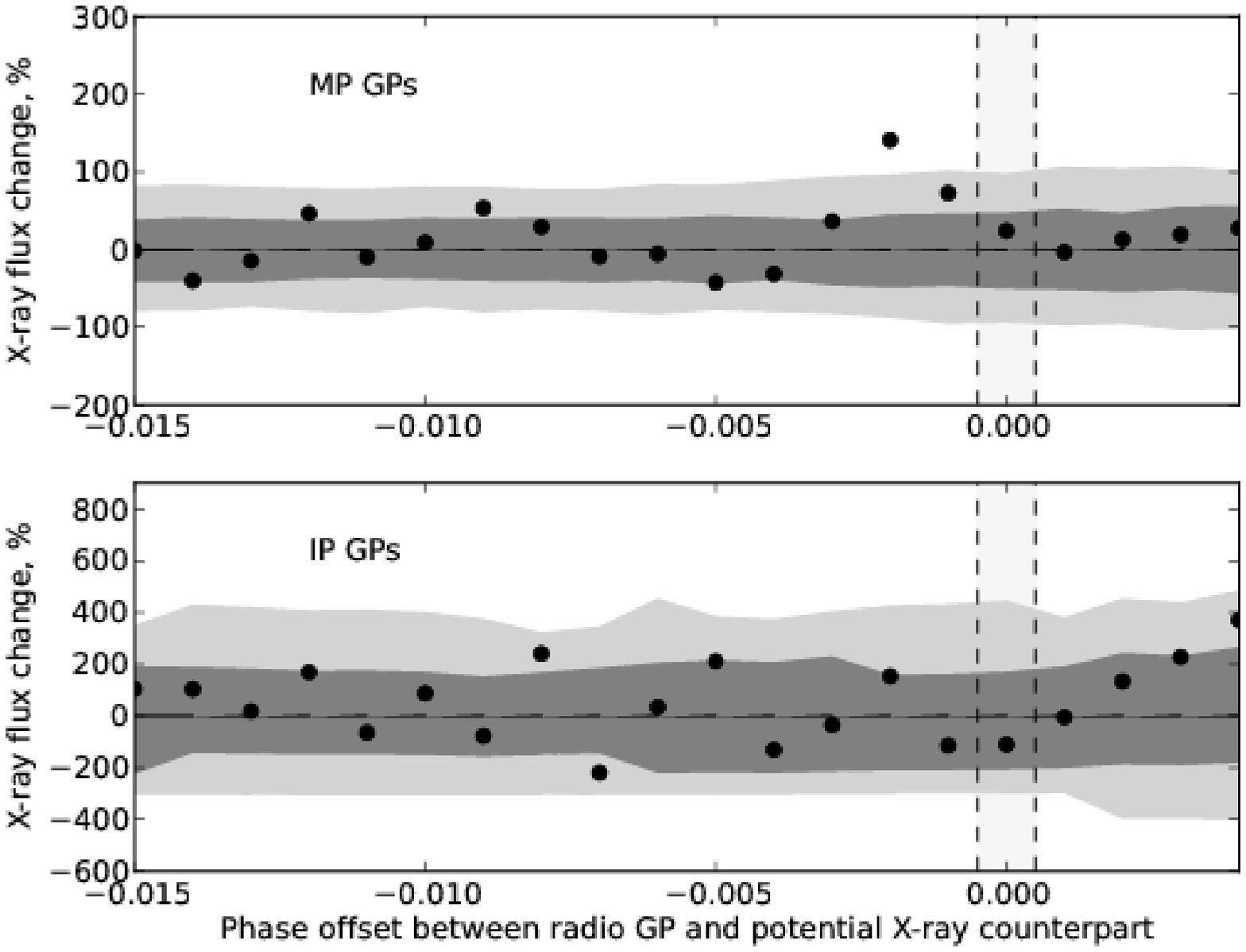}
\caption{\textit{Left:} Relative change of X-ray flux $(f_{\mathrm{GP}}-f_{\mathrm{noGP}})/f_{\mathrm{noGP}}$ in a period around a GP (within dashed lines) and in one-period windows shifted by the size of window up to $\pm15$ periods around GPs. Black dots: change of the flux around real GPs. Shaded regions -- 68\% and 95\% percentile on the pool of GP lists with random arrival times. \textit{Right}: Relative change of the X-ray flux around a GP on the timescale of the GP duration (within dashed lines) and on the same timescale but with phase lags from $-0.015$ to 0.004. For the average emission profile, the peak of the X-ray main pulse is shifted with respect to the peak of average radio pulse by $-0.01$. \label{fig:corr_e_int}} 
\end{figure*}

Ideally, for measuring the change in X-ray flux in the vicinity of GPs one should identify all GPs in the data sample and compare the X-ray flux in a selected window close to GPs ($f_{\mathrm{GP}}$) to the X-ray flux in an identical phase window close to the periods without GPs ($f_{\mathrm{noGP}}$). 

In practice, our GP samples always contain some fraction of spurious detections ($s$). We do not know the true fraction of GPs in the sample ($t$) and there is background X-ray flux ($f_{\mathrm{bkg}}$). 

Thus, if we measure the X-ray flux in some phase window close to GPs, the measured flux will be related to the true flux $f_{\mathrm{GP}}$ as follows (assuming that $s$ and $t$ are small): 

\begin{equation} 
f_{\mathrm{GP}}^{\mathrm{meas}} = f_{\mathrm{bkg}} + (1-s)f_{\mathrm{GP}} + s f_{\mathrm{noGP}}. 
\label{eq:fGPmeas}
\end{equation} 

At the same time,  by counting photons in the same window close to the randomly selected periods we can estimate the average (over all periods) X-ray flux in our correlation window: 

\begin{equation} 
f_{\mathrm{aver}}^{\mathrm{meas}} = f_{\mathrm{bkg}} + t f_{\mathrm{GP}} + (1-t)f_{\mathrm{noGP}}. 
\label{eq:favermeas}
\end{equation} 

Then, $f_{\mathrm{GP}}$ and $f_{\mathrm{noGP}}$ can be found by solving both equations simultaneously.

We performed this procedure separately for MP and IP GPs and for the three different timescales -- GP duration, radio emission phase window and pulsar period. For every timescale we used not only the correlation window centered on GPs, but also a set of windows of the same size, shifted with respect to GPs. 

 Background flux $f_{\mathrm{bkg}}$ was measured in the narrow region around the initial extraction region (Fig.~\ref{fig:xray_reg}, left, dashed line). Due to imperfect PSF modeling, we still detected weak pulsed emission in the background, so we used only photons with pulse phases 0.7$-$1.0. The fraction of false GPs $s$ was calculated as in Sec.~\ref{sec:radioobs}. For the fraction of periods with GPs, we adopted the fraction of periods with \textit{detected} GPs, 0.04 for MP GPs and 0.008 for IP GPs, although the true fraction remains unknown.  However, the limits on flux change are not sensitive to the change of $t$, as long as it remains small. For the two larger correlation timescales, the radio emission phase window and the pulsar period, varying $t$ up to 0.2 does not increase the estimated X-ray flux change in vicinity of GP by more than a factor of two.

For all correlation windows, for each GP we selected all X-ray photons within the given window (photons that satisfied the criterion for multiple GPs were counted multiple times). For comparison, we also measured $f_{\mathrm{aver}}^{\mathrm{meas}}$ in the identical windows by creating 1000 lists of fake GPs which had the same phase of arrival as real GPs, but random arrival period. We used simulated lists instead of averaging directly over all periods in order to get the errors on the X-ray flux estimates due to the limited number of photons/GPs.  The histograms of photons around real GPs were compared with mean, 68\% and 95\% percentiles on the pool of simulated datasets and the relative flux change, $(f_{\mathrm{GP}}-f_{\mathrm{noGP}})/f_{\mathrm{noGP}}$ was calculated with equations \ref{eq:fGPmeas} and \ref{eq:favermeas}. 

\textit{For all unshifted correlation windows the value of the X-ray flux stayed within the corresponding 95\% percentile measured on the pool of simulated GP lists with random arrival period.} In other words, the measured change of the X-ray flux in a window around a GP was within 2-$\sigma$ intrinsic scatter of this value due to limited number of GPs/photons. In a few shifted windows the X-ray flux exceeded 2-$\sigma$ level, but it always stayed within $3\sigma$.

\subsection{Correlation on the timescale of radio emission window}

Figure~\ref{fig:corr_e} (top part of each subplot) shows the X-ray counts in the radio emission region around MP GPs (left, within dashed lines) and IP GPs (right, within dashed lines) together with X-ray counts in the windows of same size, shifted by the size of the window within one pulsar period from a GP. The bottom part of each subplot shows the relative flux change, $(f_{\mathrm{GP}}-f_{\mathrm{noGP}})/f_{\mathrm{noGP}}$. The X-ray flux in MP/IP emission regions during periods with GPs stays within $2\sigma $ intrinsic scatter of this value due to limited number of GPs/photons, which is $\pm$10\% for MP and $\pm$30\% for IP.

\subsection{Correlation on the timescale of pulsar period}
If the process that generates GPs has somewhat less prominent long-term influence on X-ray flux, it would be easier to detect it with larger correlation windows. Figure~\ref{fig:corr_e_int} (left) shows the relative X-ray flux change for a timescale of one pulsar period around GPs (within dashed lines) and in one-period windows shifted by the size of window up to $\pm15$ pulsar periods around GPs. The X-ray flux in the same period as GPs stays within 2-$\sigma $ intrinsic scatter of this value due to limited number of GPs/photons, which is $\pm5$\% for MP and $\pm$10\% for IP. 

\subsection{Correlation on the timescale of GP duration}
\label{subsec:GP_shortest_corr}
If X-ray photons are emitted simultaneously with GPs, then the change of X-ray flux will reveal itself on timescales comparable to the widths of GP. However, if the paths of X-rays and radio emission are different throughout the magnetosphere, then the X-ray ``pulses`` will be shifted with respect to the radio GPs. Since the radio peak is delayed in phase from the X-ray one by 0.01, we looked for the correlation on the timescale of GPs with 20 phase offsets starting from $-0.015$ with a step size of 0.001.

Most of our GPs have widths less than the X-ray photon timing precision (16\,$\mic$), so only rough estimates of the change of X-ray flux was possible. We looked for the correlation in a window around each GP with the size of the window equal to the width of that GP plus 32\,$\mic$. For such choice of correlation window $s$ and $t$ depend in a complex way on the distribution of widths of the GPs. However, they bring only minor corrections, so we set both of them to 0. The change of the X-ray flux $(f_{\mathrm{GP}}-f_{\mathrm{noGP}})/f_{\mathrm{noGP}}$ was estimated from the following system of equations:  

\begin{equation} 
f_{\mathrm{GP}}^{\mathrm{meas}} \approx f_{\mathrm{bkg}} + \frac{f_{\mathrm{GP}}\cdot w_{\mathrm{GP}}}{32\mic+w_{\mathrm{GP}}} + \frac{f_{\mathrm{noGP}}\cdot32\mic }{32\mic+w_{\mathrm{GP}}}.
\end{equation}
\begin{equation}
f_{\mathrm{aver}}^{\mathrm{meas}} \approx f_{\mathrm{bkg}} + f_{\mathrm{noGP}}.
\end{equation} 

From Fig.~\ref{fig:corr_e_int} (right) shows that the relative X-ray flux on a timescale of the GP duration stays within $\pm3\sigma$ for all phase lags chosen. The X-ray flux during GPs (zero phase lag) is within $\pm2\sigma$ and does not increase more than 2 times for MP GPs and 5 times for IP GPs.

\section{Discussion} 
\label{sec:disc} 

The lack of significant change of X-ray flux during giant pulses should be accounted for by any theory of GP generation. Currently, there exist a number of giant pulse theories, which place GPs in different parts of the magnetosphere and draw in various emission mechanisms. Unfortunately, none of the theories that we are aware of quantitatively describe possible X-ray byproducts of radio GPs. Thus, below we will give only some qualitative considerations.

\citet{Gil2004} argue that GPs are due to coherent curvature emission of relativistic charged solitons, which fly along magnetic field lines in the polar region deep inside the light cylinder. The lack of strong synchrotron X-ray emission accompanying radio GPs suggests that the most particles in solitons have zero pitch angle with respect to magnetic field.

The model of \citet{Weatherall1998} explains GPs using spatial collapse of wave packets in strongly turbulent polar cap plasma. This model describes direct conversion of plasma wave turbulence into radio emission, and  since it does not imply a significant increase of particle density in the emitting volume, little correlation with high-energy incoherent photons is expected.

\citet{Harding2008} directly predicts time and phase correlation of photons below 200\,MeV with radio pulses.  Radio photons undergo resonant absorption by the secondary electron-positron plasma on the field lines interior to the slot gap. Because of this, charged particles are able to maintain significant pitch angles, and emit synchroton X-ray photons. However, quantitative limits on the energy of the photons and amount of X-ray flux change increase are not given. 

Our results rule out strong correlation between radio GPs at 1.1$-$1.9\,GHz and X-ray photons of 1.5$-$4.5\,keV. However, there can still exist weak correlation on a few percent level, similar to those observed in the optical band \citep{shearer2003} and/or correlation between GPs at some particular radio wavelength and specific high-energy band. 

% If you have acknowledgments, this puts in the proper section head. 
\bigskip 
% extra skip inserted 
\begin{acknowledgments} AVB thanks Paul Demorest (NRAO) for the help in extracting GP spectra. We thank the anonymous referee for comments that improved this paper. The National Radio Astronomy Observatory is a facility of the National Science Foundation operated under cooperative agreement by Associated Universities, Inc. This work was supported by \textit{Chandra} grant GO0-11092X. MAM is supported by a Cottrell Fellowship, a Sloan Fellowship, and a WVEPSCOR Research Challenge Grant. 
\end{acknowledgments} 
 
\bibliographystyle{apj} 
\bibliography{paper_bibliography} 

\end{document}